\documentclass[useAMS,usenatbib]{mnras}
\usepackage{natbib}
\usepackage{graphicx}
\graphicspath{ {\images} }
\usepackage{savesym}
\usepackage{listings}
\usepackage{multirow}
\usepackage[intlimits]{amsmath}
\savesymbol{iint}
\usepackage{txfonts}
\usepackage{bm}
\usepackage{amssymb}
\usepackage{amsmath}
\usepackage{tabularx}
\usepackage{color}
\usepackage{comment}
\usepackage{float}

\newcommand{\be}{\begin{equation}}
\newcommand{\ee}{\end{equation}}
\newcommand{\bea}{\begin{eqnarray}}
\newcommand{\eea}{\end{eqnarray}}

\title[Pulsar Y-point]{On the pulsar Y-point}
\author[I. Contopoulos]
       {I. Contopoulos$^1$\thanks{E-mail: icontop@academyofathens.gr}, D. Ntotsikas$^2$ and
K. N. Gourgouliatos$^2$\\
$^1$ 
Research Center for Astronomy and Applied Mathematics, Academy of Athens, Athens 11527, Greece\\
$^2$ 
Department of Physics, University of Patras, Patras, Rio, 26504, Greece
}

\begin{document}

\maketitle

\label{firstpage}

\begin{abstract}
The pulsar magnetosphere is divided into a corotating region of closed field lines 
surrounded by open field lines that emanate from the two poles of the 
star, extend to infinity and are separated by an equatorial current sheet. The three regions meet at a magnetospheric Y-point. In steady-state  solutions of the ideal force-free magnetosphere, the Y-point may lie at any distance inside the light cylinder. Time-dependent force-free simulations, however, develop closed-line regions that extend all the way to the light cylinder. On the other hand, particle (PIC) solutions consistently develop smaller closed-line regions. In order to understand this effect, we solve the pulsar equation with an improved numerical method. We show that the total electromagnetic energy stored in the ideal force-free magnetosphere manifests a subtle minimum when the closed-line region extends to only $90\%$ of the light cylinder, and thus argue that the system will spontaneously choose this particular configuration. Furthermore, we argue that the intersection of the corotating region with the equatorial current sheet is at right angles, literally leading to a T-point.
\end{abstract}


\section{Pulsar spindown and the extent of the closed-line region}

\begin{figure}
 \centering
 \includegraphics[width=8cm,angle=0.0]{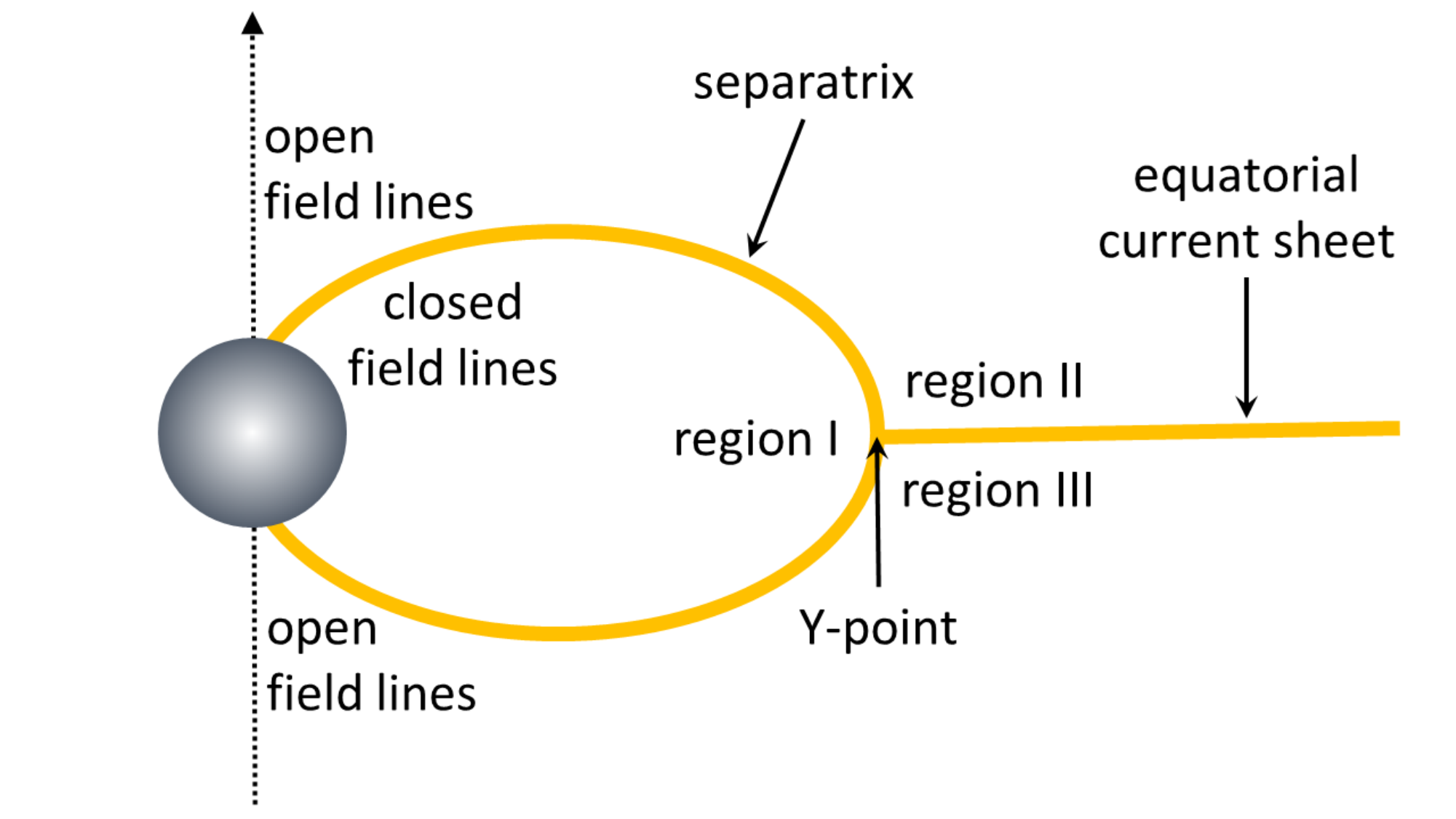}
 \caption{Schematic of magnetospheric open and closed line regions I, II, III. All three regions are separated by electric current sheets and meet at the so-called Y-point.}
\label{regions}
\end{figure}

Standard dipolar pulsar magnetospheres are divided into three regions: a region of untwisted closed field lines (hereafter region~I), and two regions of azimuthally backward twisted open field lines (hereafter regions~II and III) separated by an equatorial current sheet discontinuity (\cite{2012MNRAS.420.2793K}, \cite{2023MNRAS.518.6390S}). The closed-line region is separated from regions~II and III by a separatrix current sheet (see figure~\ref{regions}). The equatorial current sheet joins the separatrix current sheet at a singular line which in a meridional magnetospheric cross section manifests itself as a Y-point.
In the present discussion we will only consider axisymmetic magnetospheres, but our results may also be generalized for oblique rotators. 

The electromagnetic energy loss rate $L$ of the axisymmetric rotator is found numerically to be equal to
\begin{equation}
L\approx\frac{\Omega^2 \Psi_{\rm open}^2}{6\pi^2 c}\approx\frac{\Omega^2 B_*^2 r_*^6}{4 c R_{\rm Y}^2}=\frac{1}{x_{\rm Y}^2}L_{\rm canonical}
\label{L}
\end{equation}
(e.g. \cite{2005A&A...442..579C}, \cite{2006MNRAS.368.1055T}, hereafter T06, \cite{2009A&A...496..495K}). Here, $\Omega$ is the angular velocity of stellar rotation, $\Psi_{\rm open}\equiv 
\pi R_{\rm pc}^2 B_*$ is the amount of open magnetic flux that originates in the two polar caps of cylindrical radius $R_{\rm pc}\approx \sqrt{3/2}\sqrt{r_*^3/R_{\rm Y}}$ (\cite{2006ApJ...648L..51S}), $B_*$ is the polar value of the dipole magnetic field, $r_*$ is the stellar radius, $R_{\rm Y}\equiv x_{\rm Y}R_{\rm LC}$ is the distance of the Y-point beyond which magnetic field lines open up to infinity, $R_{\rm LC}\equiv c/\Omega$ is the radius of the light cylinder, and $L_{\rm canonical}\equiv \Omega^2 B_*^2 r_*^6/(4 c R_{\rm LC}^2)=\Omega^4 B_*^2 r_*^6/(4 c^3)$. In general, $x_{\rm Y}\leq 1$. Notice also that $\Omega^4 B_*^2 r_*^6/(6 c^3)$ is the electromagnetic energy loss rate of an orthogonal dipole rotator in vacuum. Eq.~(\ref{L}) is very important. It implies that the pulsar spindown rate depends strongly on the location of the Y-point. If for some reason the Y-point is located a significant distance inside the light cylinder, namely $x_{\rm Y}\ll 1$, eq.~(\ref{L}) leads to a significant overestimation of the stellar magnetic field $B_*$  (as e.g. in \cite{1999ApJ...525L.125H}). 

Steady-state  Force-Free Electrodynamic (hereafter FFE) and Magneto-Hydrodynamic (hereafter MHD) solutions of the ideal force-free magnetosphere have shown that the closed-line region is free to extend up to any distance inside the light cylinder (i.e. $x_{\rm Y}$ can have any value between $r_*/R_{\rm LC}$ and 1). Time-dependent solutions, however, always relax to a solution with the closed-line region extending as close to the light cylinder as numerically possible (as we will see next, several physical quantities diverge when the Y-point lies exactly on the light cylinder (\cite{2012ApJ...754L..12P}, \cite{2013MNRAS.435L...1T}). Over the past 10 years, a new type of numerical simulations has appeared in the literature, namely global (so-called `ab initio') PIC simulations (\cite{2014ApJ...785L..33P}, \cite{2015ApJ...815L..19P},  \cite{2015ApJ...801L..19P}). These show a consistently smaller closed-line region that extends only up to a fraction of the light cylinder radius. The extent of the closed-line region affects the pulsar spindown rate, thus, it is imperative to understand the origin of this effect. It has been theorized that this may be a numerical artifact (either the simulation has not evolved long enough to relax to a steady-state, either the inertia of the PIC particles is artificially high, either scale separation is not as large in simulations as in reality, e.g. skin depth and Larmor radii at the light cylinder vs magnetospheric size). We instead will argue in the present paper that this effect may be understood by a more physical and detailed treatment of the return current sheet in the pulsar equation. We will show that the total electromagnetic energy stored in the ideal force-free magnetosphere manifests a subtle minimum when the closed-line region extends up to $93\%$ of the light cylinder. We thus argue that the system will spontaneously choose this particular configuration which is close to the ones obtained in global PIC simulations. We will next investigate in detail the Y-point.

\section{The Y-point is in fact a T-point}

\begin{figure}
 \centering
 \includegraphics[width=8cm,angle=0.0]{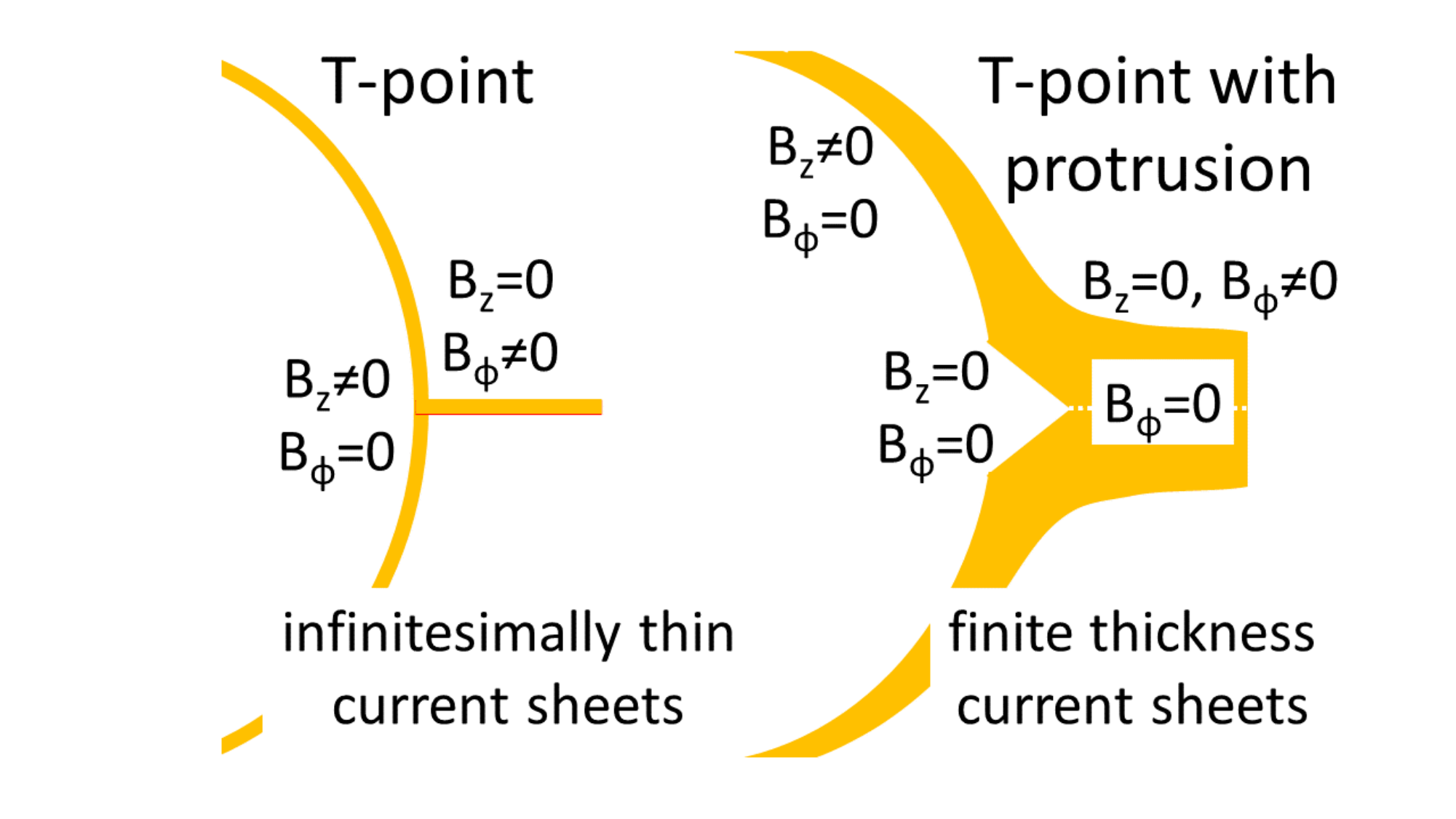}
 \caption{Schematic of idealized T-point with infinitely thin current sheets vs realistic T-point with equatorial protrusion. In the idealized T-point, $B_z$ in the closed-line region balances $B_\phi$ in the open-line region. In the realistic T-point, $B_\phi=0$ in the interior of the equatorial current sheet, and therefore, the closed-line region creates a protrusion in the equator where $B_z\rightarrow 0$.}
\label{figureT}
\end{figure}

We will be guided by \cite{2003ApJ...598..446U}~(hereafter U03), but we will also take into account what we have learned about pulsar magnetospheres over the past 20 years. We will consider only the axisymmetric case. We know today that the separatrix between open and closed field lines contains an electric current sheet which closes the global magnetospheric electric current circuit. This was not yet clear at the time of U03. This implies that the azimuthal magnetic field $B_\phi$ is non-zero right outside the Y-point, and zero inside the closed line region. Force-balance in a relativistic force-free magnetosphere implies that (\cite{1969ApJ...157..869G})
\begin{equation}
\rho_e {\bf E}+{\bf J}\times {\bf B}=0\ .
\label{forcebalance}
\end{equation}
Here, $\rho_e\equiv \nabla\cdot{\bf E}$, and ${\bf J}=\nabla\times {\bf B}$ (in steady state). U03 (see also  Lyubarky~1990) integrated eq.~(\ref{forcebalance}) accross the separatrix current sheet. This yields that
\begin{equation}
(B^2-E^2)_I=(B^2-E^2)_{II}
\end{equation}
or equivalently,
\begin{equation}
(B_p)^2_{I}=(B_p)^2_{II}+\frac{B_{\phi\ II}^2}{1-x^2}\neq 0\ ,
\label{BpI}
\end{equation}
where $B_p$ denotes the poloidal magnetic field component in each region accross the separatrix at the Y-point, $E_p\equiv x B_p$ is the poloidal component of the electric field, and $x\equiv R/R_{\rm LC}$ is the cylindrical radius in units of the radius of the light cylinder. The toroidal magnetic field component just outside the Y-point is given by
\begin{equation}
|B_{\phi}|_{II}(x_{\rm Y})=\frac{I_{\rm pc}}{2cx_{\rm Y}R_{\rm LC}}=\frac{3}{8}\frac{B_* r_*^3}{R_{\rm LC}^3 x_{\rm Y}^2}\ ,
\end{equation}
where $I_{\rm pc}\approx \rho_e \pi R_{\rm pc}^2 c =\Omega B_* R_{\rm pc}^2/2$
is the total electric current flowing though each of the pulsar polar caps. 
The magnetic field in region~I must obey the pulsar equation without poloidal electric current, namely,
\begin{equation}
(1-x^2)\left(\frac{\partial^2\Psi}{\partial x^2}+\frac{\partial^2\Psi}{\partial z^2}\right)-\frac{1+x^2}{x}\frac{\partial\Psi}{\partial x}=0\ .
\end{equation}
Spatial cylindrical coordinates $x$ and $z$ are expressed here in units of the light cylinder radius $R_{\rm LC}$. The magnetic field components can be written in terms of the magnetic flux function $\Psi(x,z)$ as
\begin{equation}
(B_R)_{I}=-\frac{1}{2\pi R_{\rm LC}^2}\frac{1}{x}\frac{\partial\Psi}{\partial z}\ ,\ \\
(B_z)_{I}=\frac{1}{2\pi R_{\rm LC}^2}\frac{1}{x}\frac{\partial\Psi}{\partial x}\ ,\ \\
(B_\phi)_{I}=0\ 
\end{equation}
Following U03, we will introduce polar coordinates $(r,\theta)$ around the Y-point, such that
\begin{equation}
x=x_{\rm Y}-r\cos\theta\ ,\\ z=r\sin\theta\ .
\end{equation}
In these coordinates, we can rewrite the pulsar equation in region $I$ as
\begin{eqnarray}
&& (1-x_{\rm Y}^2)\left(\frac{\partial^2\Psi}{\partial r^2}+\frac{1}{r}\frac{\partial\Psi}{\partial r}+\frac{1}{r^2}
\frac{\partial^2\Psi}{\partial \theta^2}\right)\nonumber
\end{eqnarray}
\begin{equation}-\frac{1+x_{\rm Y}^2}{x_{\rm Y}}\left(\cos\theta \frac{\partial\Psi}{\partial r}-\frac{\sin\theta}{r}\frac{\partial\Psi}{\partial \theta}\right)=0\ .
\label{pulsar2}
\end{equation}
very close to the Y-point, we will make the self-similar Ansatz that
\begin{equation}
\Psi_{I}\equiv r^\alpha f(\theta)
\end{equation}
and therefore,
\begin{equation}
(B_r)_{I}=\frac{1}{2\pi R_{\rm LC}^2}r^{\alpha-1}f'(\theta)\ ,\\
(B_\theta)_{I}=-\frac{1}{2\pi R_{\rm LC}^2}\alpha r^{\alpha-1}f(\theta)\ .
\end{equation}
We will also assume that $\Psi=0$ along the separatrix.
Obviously, in order for $B_p$ to be finite in region~I all the way down to $r\rightarrow 0$ as required by eq.~(\ref{BpI}), $\alpha$ must be equal to 1. In the limit $r\rightarrow 0$, eq.~(\ref{pulsar2}) then becomes
\begin{equation}
f''(\theta)=f(\theta)\ .
\end{equation}
Since at $\theta=\pi$ the field crosses the equator vertically, and thus $B_r(\theta=\pi)=0$, this yields that $f'(\theta=\pi)=0$, which yields the solution
\begin{equation}
f(\theta)\propto \cos(\theta)
\end{equation}
Obviously, $\Psi=0$ where $f=0$, thus the separatrix lies at
\begin{equation}
\theta_s=\pi/2\ .
\end{equation}
In other words, technically, the Y-point is a T-point as can be seen in figure~\ref{figureT}\footnote{Notice that \cite{2005PhRvL..94b1101G} assumed instead that $\alpha=0.5$, from which he derived $\theta_s=77.3^\circ$.}. Obviously, just outside a T-point,
\begin{equation}
(B_p^2)_{II}(x_{\rm Y})=0
\end{equation}
and eq.~(\ref{BpI}) then yields that
\begin{equation}
|B_p|_{I}(x_{\rm Y})=\frac{|B_{\phi}|_{II}(x_{\rm Y})}{\sqrt{1-x_{\rm Y}^2}}=\frac{3 B_*}{8}\frac{r_*^3}{R_{\rm LC}^3}\frac{1}{x_{\rm Y}^2\sqrt{1-x_{\rm Y}^2}}\ .
\label{BpY}
\end{equation}

\begin{figure}
 \centering
 \includegraphics[width=8cm,angle=0.0]{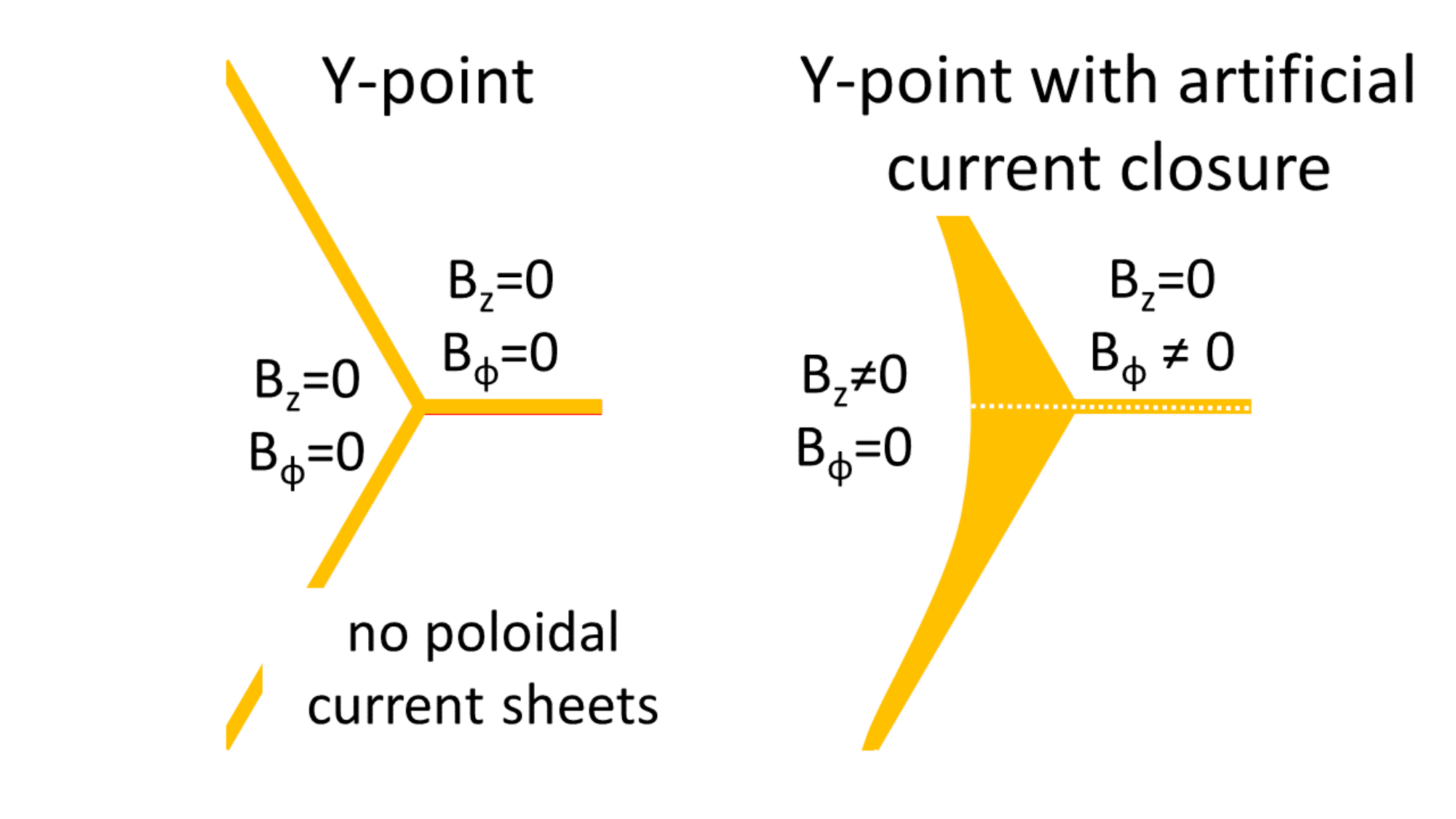}
 \caption{Schematic of idealized Y-point with infinitely thin current sheets vs Y-point with current closure in the closed-line region as in CKF. The first one is viable only if there is no jump in $B_\phi$ accross the separatrix  (i.e. $B_\phi=0$ or constant everywhere around the Y-point), thus it does not apply to the pulsar magnetosphere as we understand it today. The second one is unrealistic (current closure must take place along open field lines, not along closed lines). The field crosses the equator vertically inside the current sheet, and at a non-vertical angle outside.}
\label{figureY}
\end{figure}

Global PIC simulations of \cite{2022ApJ...939...42H}, and \cite{2023ApJ...943..105H} show a different picture around the Y-point. In particular, instead of it being a T-point, on the contrary, it seems to locally protrude outwards. We believe the answer is that due to the presence of an electric current sheet immediately outside the closed-line region, there is a point exactly along the equator where $B_\phi=0$. The above analysis obviously brakes down around that point, and the closed line region protrudes outwards like a `hernia'. 
The height of the protruding region is equal to the thickness of the equatorial current sheet. 
This effect is seen clearly in the high-resolution PIC simulations of \cite{2022ApJ...939...42H} which resolve in detail the equatorial current sheet. It is interesting that this effect has been seen before in \cite{10.1093/mnras/sty2766} (figure~2 top left) and \cite{2023IAUS..363..338N}. There is a simple explanation why only these solutions of the pulsar equation show this effect: these are the only solutions of the pulsar equation known in the literature where the return current is placed {\it inside} the last open field line $\Psi=\Psi_{\rm open}$, in the regions of open field lines II and III. 

When solving the pulsar equation, the distribution of the poloidal current 
along the open magnetic field lines is determined from the condition of smooth crossing of the light cylinder singular surface (\cite{1999ApJ...511..351C}, hereafter CKF).  This procedure, however, does not take into account the return current along the separatrix which must be specifically dealt with. Mathematically, the return current corresponds to an infinitely abrupt jump of the magnetospheric electric current $I(\Psi)$ from $I_{\rm pc}$ to zero. In practice, it may be viewed as half a Gaussian distribution of height $I_{\rm pc}$ and width $\delta\Psi\ll\Psi_{\rm open}$. In that narrow region, the force-free conditions implied by the pulsar equation do not apply, thus this narrow region is problematic in the context of the pulsar equation. 
CKF first implemented a current distribution along closed lines $\Psi_{\rm open}\leq\Psi\leq\Psi_{\rm open}+\delta\Psi$. Obviously, the specified electric current does not cross the light cylinder. It is only a mathematical approximation of the return current distribution that allows us to solve the pulsar equation. \cite{2005PhRvL..94b1101G} and T06 followed a similar approach.
The CKF prescription guaranteed that the last closed line without poloidal electric current crosses the equator vertically (see. e.g. figure~4 of T06 and the right panel in figure~\ref{figureY}) and does not form the protrusion observed in recent simulations. \cite{10.1093/mnras/sty2766} were the first to place it {\it inside} the open line region, 
\begin{equation}
\Psi_{\rm open}-\delta\Psi\leq \Psi\leq\Psi_{\rm open}\ .
\label{return}
\end{equation}
In these solutions, the poloidal electric current right outside the Y-point on the equator is equal to zero, hence the equatorial field protrusion has nothing to do with inertia. In fact, the region of open poloidal flux over which the return current sheet flows is equal to $\delta\Psi\approx 0.01\Psi_{\rm open}$ in \cite{10.1093/mnras/sty2766} and  $\delta\Psi=0.005\Psi_{\rm open}$ in \cite{2023IAUS..363..338N}, hence the corresponding protrusions are correspondingly thinner. As we have spread the current sheet into a narrow layer, we find that the regularisation condition is not completely fulfilled in this region. Because of this, some features just outside the light cylinder may appear, in the form of magnetic islands. These become negligible for they affect the solution only in a thin layer i.e.~one that corresponds to $\delta \Psi=0.005\Psi_0$, but may become significant for solutions with larger $\delta \Psi$.

\begin{figure}
 \centering
 \subsection*{a}
 \includegraphics[width=7cm,angle=0.0]{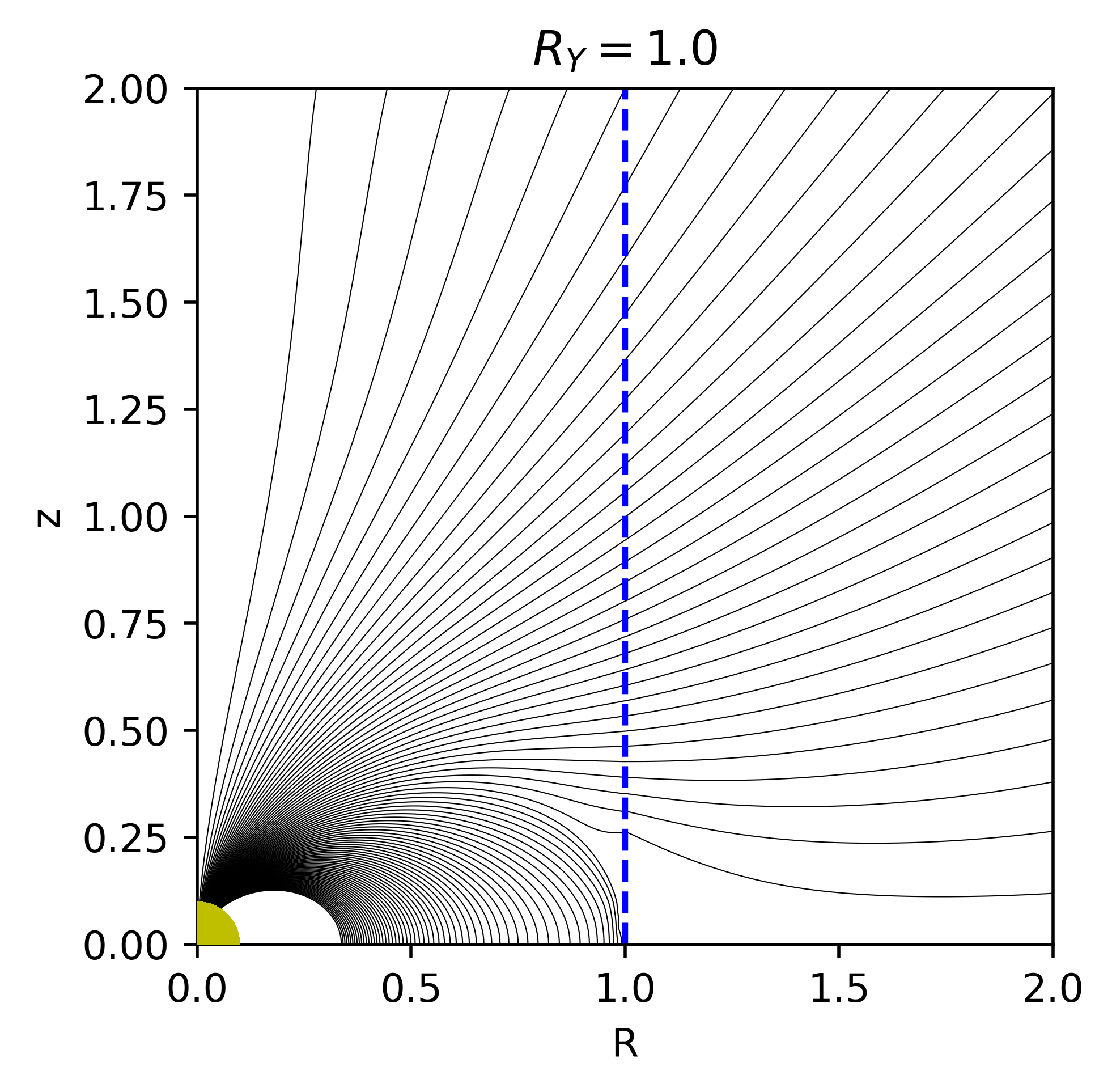}
 \vspace{-0.8cm}
 \subsection*{b}
 \includegraphics[width=7cm,angle=0.0]{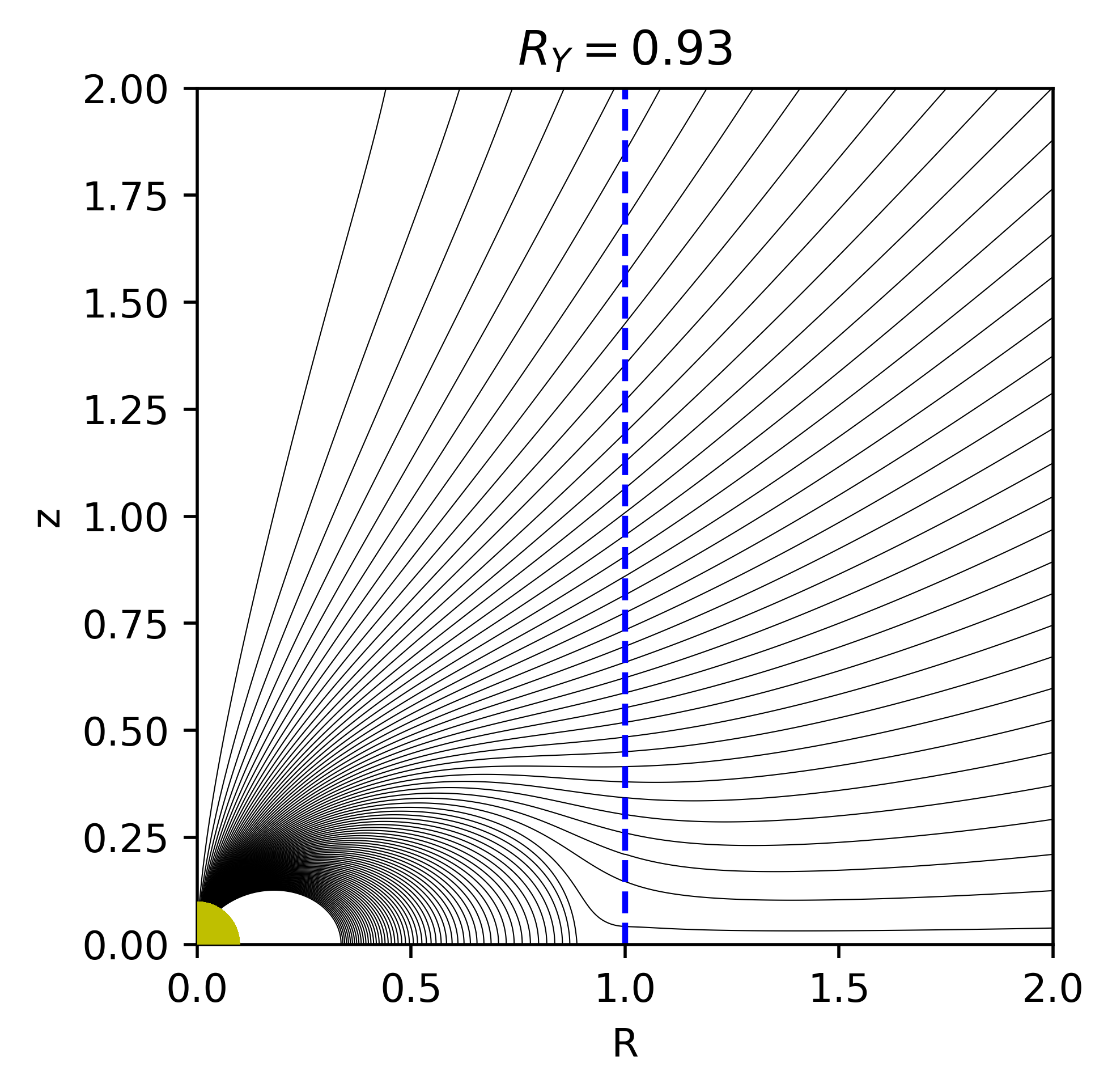}
 \vspace{-0.8cm}
 \subsection*{c}
 \includegraphics[width=7cm,angle=0.0]{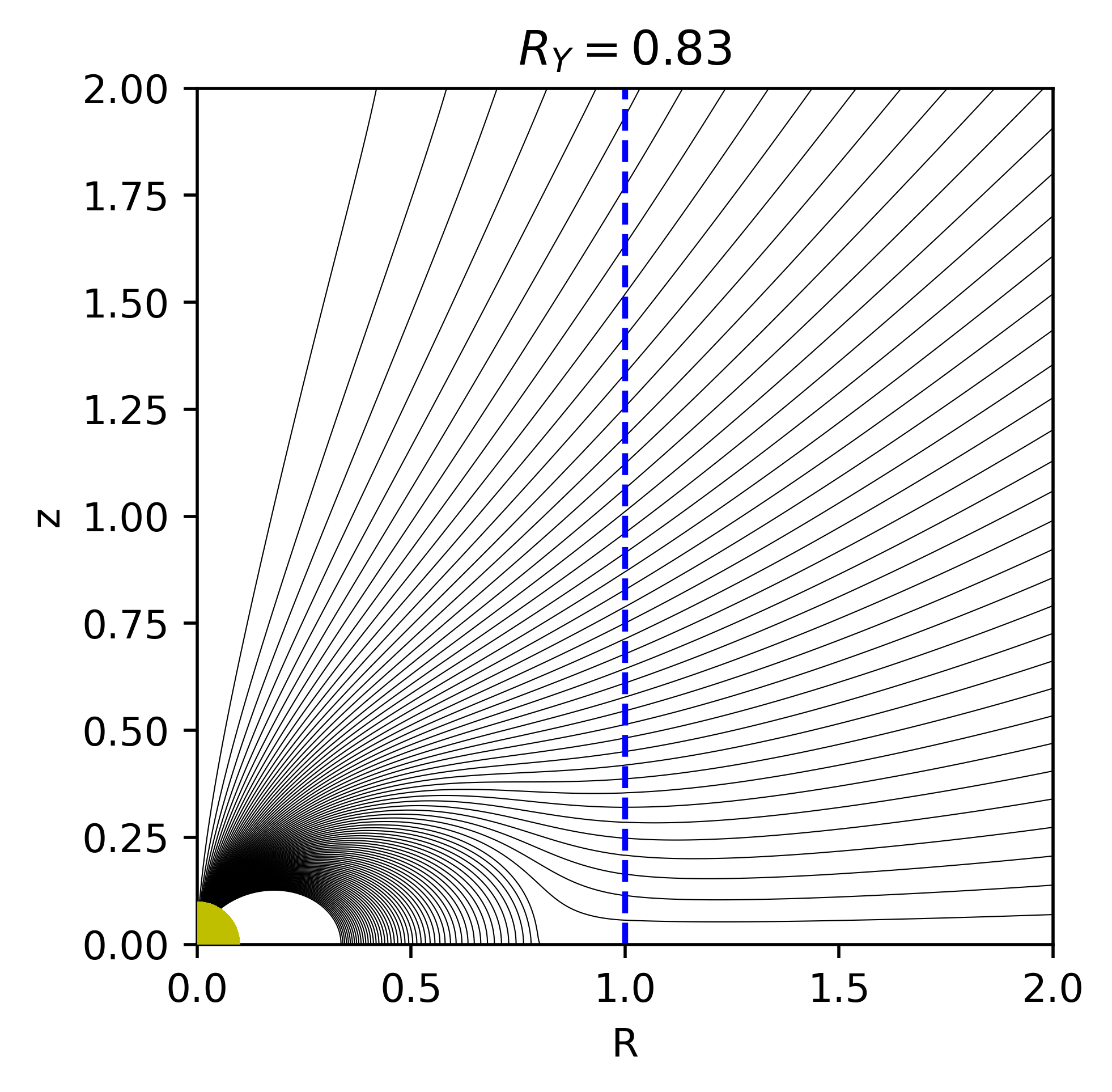}
 \caption{High-resolution solutions for various values of $x_{\rm Y}\leq 1$. The return current is imposed to flow along the last open field lines in regions II \& III. It is clearly seen that the Y-point is in fact a T-point.}
\label{solutions}
\end{figure}

We obtained new high resolution solutions of the pulsar equation with the return current imposed over the last open field lines above the separatrix between open and closed field lines. We have used an elliptic solver (utilizing the Successive Overrelaxation Method) in the computational domain $0<x<2$ and $0<z<2$ with a resolution of 800 points in $x$ equally spaced inside the light cylinder, 800 points in $x$ equally spaced outside the light cylinder, and 800 equally space points in $z$  and we find that the solution converges after $2\times10^8$ iterations  for the cases of simulations where $R_Y =0.93, R_Y = 0.83$ (in principle for any simulation where the current sheet is placed at $R_Y \le 0.98$), while for the simulation with the current sheet placed at $R_Y = 0.99$ and $R_Y = 1.0 $ the solution converges after $10^9$ iterations. Our resolution is higher than CKF but lower than T06. The distribution of the magnetospheric electric current $I(\Psi)$ was iteratively adjusted by the condition of smooth crossing of the light cylinder as described in \cite{10.1093/mnras/sty2766}. To account for the return current flowing on the separatrix between the open closed field lines, we approximated the $\delta$-function return current in the open-line region of eq.~(\ref{return}) by a narrow Gaussian of height $I_{\rm pc}$ and width $\delta\Psi=0.005 \Psi_{\rm open}$ centered at $\Psi=\Psi_{\rm open}$. We found that indeed, the Y-point is clearly a T-point for all values of $x_{\rm Y}$ (figure~\ref{solutions}), unlike the solutions shown in figure 4 of T06 which develop clear Y-points. We understand this discrepancy with the right panel of figure~\ref{figureY}, where a separatrix without a jump in $B_\phi$ crosses the equator at a nonzero angle, while the innermost closed field line where the imposed return current flows crosses the equator vertically. Unfortunately, as we argued above, such a current closure configuration is unphysical.
It is interesting that time-dependent force-free and PIC runs yield oblique Y-points, not T-points. We suspect that all such runs contain a very thick current sheet where what we superficially observe as a Y-point is in fact the thick and extended protrusion shown schematically in the right panel of figure~2.

\section{A subtle energy minimum}

According to \cite{2005A&A...442..579C} and T06, the Y-point can lie anywhere inside the light cylinder (it can certainly not lie outside). This effect has been corroborated by a study of the total magnetospheric energy content as a function of the position of the Y-point (see figure~10 of T06). Unfortunately, the analysis of what happens around the Y-point is subtle, and requires a more careful numerical treatment with high resolution.

\begin{figure}
 \centering
 \includegraphics[width=9cm,angle=0.0]{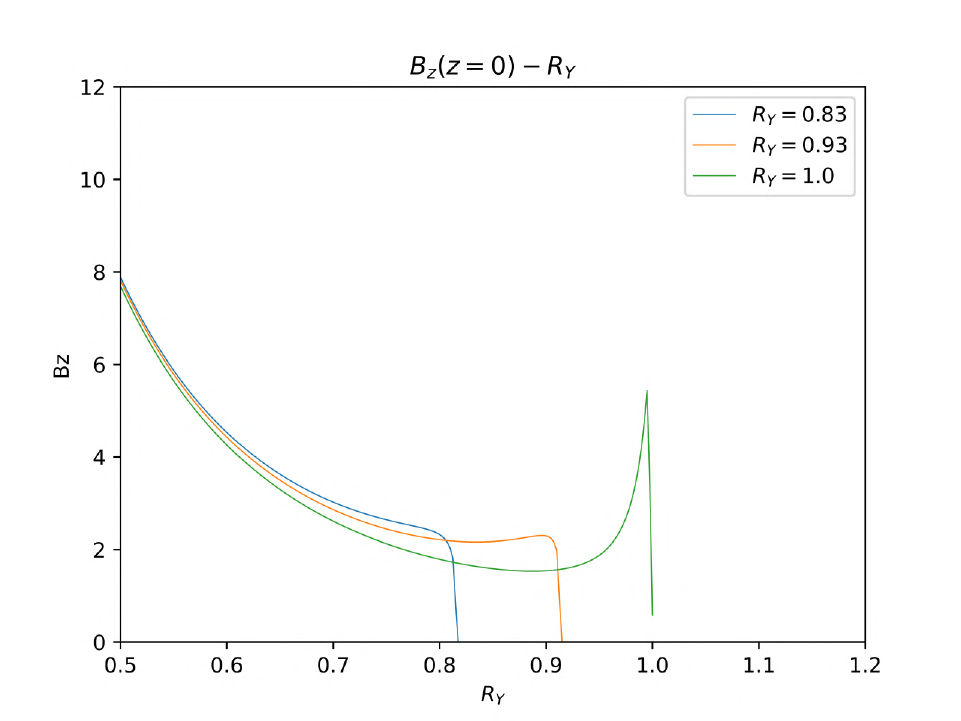}
 \caption{Poloidal magnetic field strength in the equatorial plane as a function of $R/R_{\rm LC}$ for the high-resolution solutions shown in figure~\ref{solutions}. $B_z$ is normalized to $B_* r_*^3/R_{\rm LC}^3$. The divergence of $Bz$ near the Y-point is much more pronounced than in all previous solutions in the literature.}
\label{figureBz}
\end{figure}

Eq.~(\ref{BpY}) tells us that $|B_p(x_{\rm Y})|_I$ decreases $\propto 1/x_{\rm Y}^2$ as the Y-point is moved outwards, but beyond some distance it increases again as $x_{\rm Y}\rightarrow 1$ (see figure~\ref{figureBz}). This makes us suspicious that indeed, the electromagnetic energy of the magnetosphere increases as $x_{\rm Y}$ moves beyond some distance and some part of the open line region that contains normal valued $B_p$ and $B_\phi$ is replaced by a local region of enhanced poloidal magnetic field $\propto 1/\sqrt{1-x^2}$. We performed this detailed calculation and found a subtle local minimum of the integral
\begin{equation}
W\equiv \int_{r_*}^{2R_{\rm LC}}\int_{0}^{2R_{\rm LC}}(B^2+E^2)R_{\rm LC}^3\ 2\pi x\ {\rm d}x\ {\rm d}z
\label{energy}
\end{equation}
for $x_{\rm Y}=0.93$. Notice that we have arbitrarily chosen an inner boundary of $x_{\rm in}=r_*\equiv 0.1R_{\rm LC}$ and $z_{\rm in}=0$, and an outer boundary of $x_{\rm out}=z_{\rm out}=2R_{\rm LC}$. The local energy minimum is very subtle because it requires a detailed high resolution treatment of the region around the Y-point when the latter approaches the light cylinder. The highest resolution to-date solution of the pulsar equation (T06) missed this effect because it explicitly did not include the equatorial region around $z=0$ in the energy integral where most of the increase in eq.~(\ref{BpY}) takes place. 

One further reason is the introduction of the separatrix return current {\it inside} the closed-line region, whereas in reality it flows {\it outside}. This artificial effect essentially removes from the calculation of the energy integral the interesting region adjacent to the separatrix where the poloidal magnetic field of the closed-line region increases dramatically. Without a detailed treatment of the region around the Y-point, the electromagnetic energy integral in eq.~(\ref{energy}) is found to be a decreasing function of $x_{\rm Y}$, and it is therefore natural to conclude that the pulsar magnetosphere will attain the minimum energy configuration that corresponds to its maximum $x_{\rm Y}$ value, namely $x_{\rm Y}=1$. 
When the region around the Y-point is considered more carefully, as in the present paper, the poloidal field divergence at the tip of the closed-line region becomes much more dramatic. Let us calculate here the energy integral of eq.~(\ref{energy}) at the tip of the closed-line region inside the Y-point. This yields
\begin{equation}
W_{\rm Y}\equiv \int^{x_{\rm Y}}\int_{z=-h(x)}^{h(x)}(B^2+E^2)\ 2\pi R_{\rm LC}^3 x\ {\rm d}x\ {\rm d}z\ ,
\label{energy1}
\end{equation}
where, $h(x)$ is the height of the tip of the closed-line region as $x\rightarrow x_{\rm Y}$. If the tip of the closed line region is a Y-point at some non-vertical angle $\theta_{\rm Y}$ (e.g. $\theta_{\rm Y}=77.3^\circ$ as calculated by \cite{2005PhRvL..94b1101G}, then $h(x)=(x_{\rm Y}-x)\tan(\theta_{\rm Y})$, whereas if it is a T-point as first shown by U03, then $h(x)=h_{\rm Y}=$~constant. Therefore, if $B(x)\sim E(x)\sim B_o /\sqrt{1-x}$ as $x\rightarrow x_{\rm Y}\rightarrow 1$, eq.~(\ref{energy1}) yields
\begin{eqnarray}
W_{\rm Y}&=& 8\pi R_{\rm LC}^3 B_o \tan(\theta_{\rm Y})\ \delta x=\mbox{finite}\ \mbox{for a Y-point}\ ,\nonumber\\
&\approx& -8\pi R_{\rm LC}^3 B_o h_{\rm Y}\ln(1-x_{\rm Y})|^1=\mbox{infinite}\ \mbox{for a T-point}\ .\nonumber\\
\label{energy2}
\end{eqnarray}
In other words, the electromagnetic energy contained in the tip of the closed-line region diverges due to the divergence of $B_z(x\rightarrow 1-,z=0)$. This is the reason the Y-point must lie at a finite region inside the light cylinder. Nevertheless, while the global energy argument is certainly interesting, it is not clear to us what would keep the Y-point from moving towards the light cylinder via field line reconnection. We suspect that, even if such field reconnection takes place, it will be locally favorable to form and eject plasmoids from the Y-point as seen in the \cite{2022ApJ...939...42H} numerical simulations. Plasmoid formation at the Y-point for various positions of the Y-point needs further investigation.

\section{Conclusions}

In this short letter we corrected some common misconseptions about the shape and the position of the magnetospheric Y-point. We showed that the pulsar magnetosphere manifests a subtle global electromagnetic energy minimum when its closed-line region ends at about $90\%$ of the light cylinder distance. This explains a result seen in all global PIC numerical simulations of the past decade. This subtle modification of the pulsar magnetosphere does not affect significantly its main properties, namely its electromagnetic energy loss and the resulting pulsar spin down rate. It also does not explain the divergence of the pulsar braking index $n$ from its canonical dipolar field value (according to eq.~\ref{L}, for a fixed value of $x_{\rm Y}$, the electromagnetic energy loss rate remains proportional to $\Omega^4$, hence $n=3$). 

\begin{figure}
 \centering
 \includegraphics[width=9cm,angle=0.0]{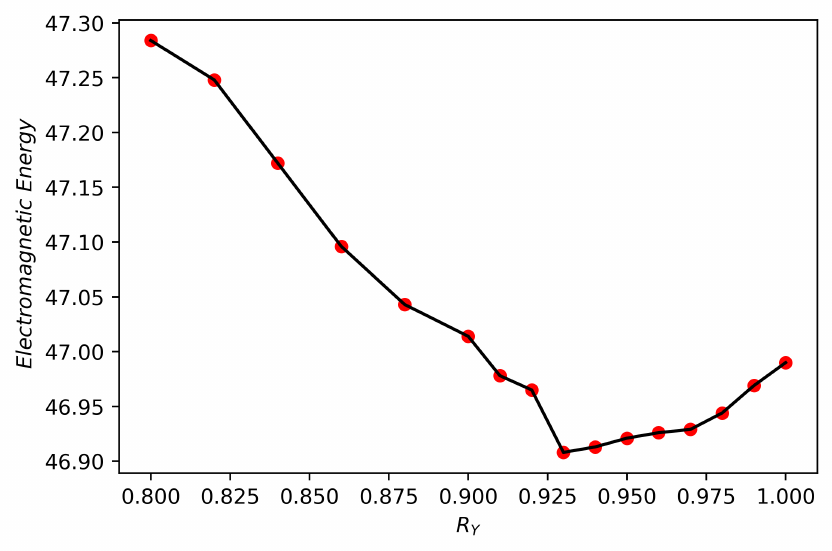}
 \caption{Electromagnetic energy inside $2R_{\rm LC}$ as a function of the extent $x_{\rm Y}$ of the closed-line region. We obtain a subtle minimum around $x_{\rm Y}\sim 0.93$.}
\label{figureEnergy}
\end{figure}

\section*{Acknowledgements}

We would like to thank the International Space Science Institute (ISSI) for providing financial support for the organization of the meeting of ISSI Team No 459 led by I. Contopoulos and D. Kazanas where the issues addressed in this work were first discussed.

\section*{Data availability statement}
The data underlying this article will be shared on reasonable request to the corresponding author.
\bibliographystyle{mnras}
\bibliography{YPoint.bib}
\label{lastpage}

\end{document}